\documentclass[12pt,a4paper]{article}
\usepackage{epsfig}
\begin{document}
\title{Lepton flavor violating processes $l_{i}\longrightarrow l_{j}\nu_{l}\bar{\nu}_{l}$ in
topcolor-assisted technicolor models }
\author{Chongxing Yue$^{a}$ and Lanjun Liu$^{b}$\\
{\small a: Department of Physics, Liaoning  Normal University,
Dalian 116029. P.R.China}\thanks{E-mail:cxyue@lnnu.edu.cn}
\\ {\small b:College of Physics and Information Engineering,}\\
\small{Henan Normal University, Xinxiang  453002. P.R.China}
 }
\date{\today}
\maketitle
\begin{abstract}
\hspace{5mm}We study the lepton flavor violating (LFV) processes
$l_{i}\longrightarrow l_{j}\nu_{l}\bar{\nu}_{l}$ in the context of
the topcolor-assisted technicolor (TC2) models. We find that the
branching ratios $B_{r}(\tau\longrightarrow
l_{j}\nu_{\tau}\bar{\nu}_{\tau})$ are larger than the branching
ratios $B_{r}(\tau\longrightarrow l_{j}\nu_{l}\bar{\nu}_{l})$ in
all of the parameter space. Over a wide range of parameter space,
we have $B_{r}(\tau\longrightarrow
l_{j}\nu_{\tau}\bar{\nu}_{\tau})\sim 10^{-6}$ and
$B_{r}(\tau\longrightarrow l_{j}\nu_{l}\bar{\nu}_{l})\sim 10^{-9}
(l=\mu$ or $e)$. Taking into account the bounds given by the
experimental upper limit
$Br^{exp}(\mu\longrightarrow3e)\leq1\times10^{-12}$ on the free
parameters of TC2 models, we further give the upper limits of the
LFV processes $l_{i}\longrightarrow l_{j}\nu_{l}\bar{\nu}_{l}$. We
hope that the results may be useful to partly explain the data of
the neutrino oscillations and the future neutrino experimental
data might be used to test TC2 models.
\end {abstract}

\newpage
It is well known that the individual lepton numbers $L_{e}$,
$L_{\mu}$ and $L_{\tau}$ are automatically conserved and the
tree-level lepton flavor violating (LFV) processes are absent in
the standard model (SM). However, the solar neutrino experiments
\cite{y1}, the data on atmospheric neutrinos obtained by the
Super-Kamiokande Collaboration \cite{y2}, and the results from the
Kam LAND reactor antineutrino experiments \cite{y3} provide very
strong evidence for mixing and oscillation of the flavor
neutrinos, which imply that the separated lepton number are not
conserved. Thus, the SM requires some modification to account for
the pattern of neutrino mixing suggested by the data and the LFV
processes like $l_{i}\longrightarrow l_{j}\gamma$ and
$l_{i}\longrightarrow l_{j}l_{k}l_{l}$ are allowed. The
observation of these LFV processes would be a clear signature of
new physics beyond the SM, which has been widely studied in
different scenarios such as Two Higgs Doublet Models,
Supersymmetry, Grand Unification \cite{y4,y5} and topcolor models
\cite{y6}.

On the other hand, neutrino oscillations imply that there are
solar $\nu_{e}\longrightarrow\nu_{\mu}, \nu_{\tau}$ transitions
and there are atmospheric $\nu_{\mu}\longrightarrow\nu_{\tau}$
transitions. The standard tau decays
$\tau\longrightarrow\mu\nu_{\mu}\bar{\nu}_{\tau}$,
$e\nu_{e}\overline{\nu}_{\tau}$ and the standard muon decay
$\mu\longrightarrow e\nu_{e}\bar{\nu}_{\mu}$ can not explain the
experimental fact. However, the LFV processes
$l_{i}\longrightarrow l_{j}\nu_{l}\bar{\nu}_{l}$, where
$l_{i}=\tau$ $or$ $\mu$, $l_{j}=\mu$ $or$ $e$ and $l=\tau, \mu$
$or$ $e$, might explain the neutrino oscillation data. With these
motivations in mind, we study the LFV processes
$l_{i}\longrightarrow l_{j}\nu_{l}\bar{\nu}_{l}$ in the context of
topcolor-assisted technicolor (TC2) models  \cite{y7}. These
models predict the existence of the extra $U(1)$ gauge boson $Z'$,
which can induce the tree-level FC coupling vertices
$Z'l_{i}l_{j}$. The effects of the gauge boson $Z'$ on the LFV
processes $l_{i}\longrightarrow l_{j}\gamma$,
$l_{i}\longrightarrow l_{j}l_{k}l_{l}$ and $Z\longrightarrow
l_{i}l_{j}$ have been studied in Ref.\cite{y6,y8}. They have shown
that the contributions of $Z'$ to these processes are
significantly large, which may be detected in the future
experiments. In this letter, we show that the $Z'$ can generate
large contributions to the LFV processes
$\tau\longrightarrow\mu\nu_{\tau}\bar{\nu}_{\tau}$,
$e\nu_{\tau}\bar{\nu}_{\tau}$ and $\mu\longrightarrow
e\nu_{\tau}\bar{\nu}_{\tau}$, which may be used to partly explain
the data of the neutrino oscillations. Furthermore, considering
the constraints of the present experimental bound on the LFV
process $\mu\longrightarrow3e$ on the free parameters of TC2
models, we give the upper bounds on the branching ratios
$Br(l_{i}\longrightarrow l_{j}\nu_{l}\bar{\nu}_{l})$, which arise
from $Z'$ exchange.

For TC2 models, the underlying interactions, topcolor interaction,
are non-universal. This is an essential feature of TC2 models, due
to the need to single out the top quark for condensate. Therefore,
TC2 models predict the existence of the non-universal $U(1)$ gauge
boson $Z'$. The new particle treats the third generation fermions
differently from those in the first and second generations and can
lead to the tree-level FC couplings. The flavor-diagonal couplings
of $Z'$ to leptons  can be written as \cite{y7,y9}:
\begin{eqnarray}
{\cal L}^{FD}_{Z'}=&-&\frac{1}{2}g_{1}cot\theta'
Z'_{\mu}(\bar{\tau}_{L}\gamma^{\mu}\tau_{L}+2\bar{\tau}_{R}\gamma^{\mu}\tau_{R}
-\bar{\nu}_{\tau L}\gamma^{\mu}\nu_{\tau L})-\frac{1}{2}g_{1}
tan\theta'Z'_{\mu}(\bar{\mu}_{L}\gamma^{\mu}\mu_{L}\nonumber\\
&+&2\bar{\mu}_{R}\gamma^{\mu}\mu_{R}+\bar{\nu}_{\mu
L}\gamma^{\mu}\nu_{\mu L}+\bar{e}_{L}\gamma^{\mu}e_{L}+
+2\bar{e}_{R}\gamma^{\mu}e_{R}+\bar{\nu}_{eL}\gamma^{\mu}\nu_{eL}),
\end{eqnarray}
where $g_{1}$ is the ordinary hypercharge gauge coupling constant,
$\theta'$ is the mixing angle with $tan\theta'=g_{1}/\sqrt{4\pi k_{1}}$.
The flavor changing couplings of $Z'$ to leptons can be written as:
\begin{eqnarray}
{\cal
L}^{FC}_{Z'}=&-&\frac{1}{2}g_{1}Z'_{\mu}[k_{\tau\mu}(\bar{\tau}_{L}\gamma^{\mu}\mu_{L}
+2\bar{\tau}_{R}\gamma^{\mu}\mu_{R})+k_{\tau
e}(\bar{\tau}_{L}\gamma^{\mu}e_{L}
+2\bar{\tau}_{R}\gamma^{\mu}e_{R})\nonumber\\
&+&k_{\mu
e}tan^{2}\theta'(\bar{\mu}_{L}\gamma^{\mu}e_{L}+2\bar{\mu}_{R}\gamma^{\mu}e_{R})],
\end{eqnarray}
where $k_{ij}$ are the flavor mixing factors. For the sake of
simplicity, we consider the case where all three generations of
leptons mix with a universal constant $k$, i.e.
$k_{\tau\mu}=k_{\tau e}=k_{\mu e}=k$ in this letter.

\begin{figure}[hb]
\begin{center}
\begin{picture}(200,100)(0,0)
\put(-80,-400){\epsfxsize130mm\epsfbox{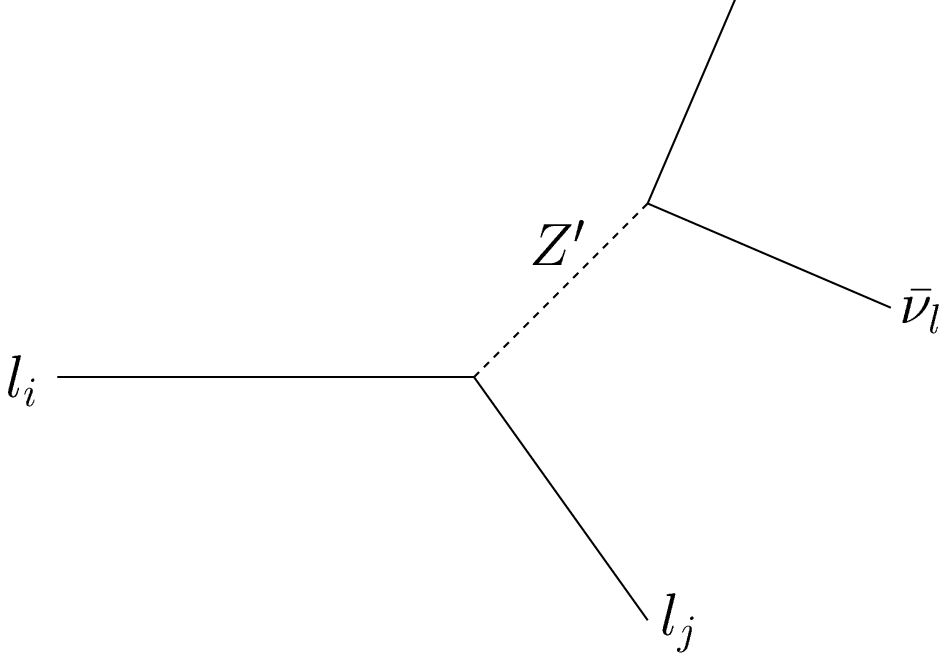}}
\put(-120,-70){Fig.1\hspace{5mm}Feynman diagram for the LFV
processes $l_{i}\longrightarrow l_{j}\nu_{l}\bar{\nu}_{l}$ induced
by $Z'$ exchange.}
\end{picture}
\end{center}
\end{figure}

\vspace{2.5cm}

From Eq.(1) and Eq.(2), one can see that the LFV processes
$l_{i}\longrightarrow l_{j}\nu_{l}\bar{\nu}_{l}$ can be generated via gauge boson $Z'$ exchange
at tree-level. The relevant Feynman diagrams are depicted in Fig.1. The partial widths can be
written as:
$$\Gamma_{1}=\Gamma(\tau\rightarrow\mu\nu_{\tau}\bar{\nu}_{\tau})=\Gamma(\tau\rightarrow
e\nu_{\tau}\bar{\nu}_{\tau})=\frac{5k_{1}\alpha_{e}}{384\pi C_{W}^{2}}\frac{m_{\tau}^{5}}{M_{Z'}^{4}}k^2,$$
$$\Gamma_{2}=\Gamma(\tau\rightarrow\mu\nu_{\mu}\bar{\nu}_{\mu})=\Gamma(\tau\rightarrow
\mu\nu_{e}\bar{\nu}_{e})=\Gamma(\tau\rightarrow
e\nu_{\mu}\bar{\nu}_{\mu})=\Gamma(\tau\rightarrow
e\nu_{e}\bar{\nu}_{e})=\frac{5\alpha_{e}^3}{384\pi k_{1}C_{W}^{6}}\frac{m_{\tau}^{5}}{M_{Z'}^{4}}k^2,$$
$$\Gamma_{3}=\Gamma(\mu\rightarrow e\nu_{\tau}\bar{\nu}_{\tau})=\frac{5\alpha_{e}^3}{384\pi k_{1}
C_{W}^{6}}\frac{m_{\mu}^{5}}{M_{Z'}^{4}}k^2,$$
$$\Gamma_{4}=\Gamma(\mu\rightarrow e\nu_{\mu}\bar{\nu}_{\mu})=\Gamma(\mu\rightarrow e
\nu_{e}\bar{\nu}_{e})=\frac{5\alpha_{e}^5}{384\pi
k_{1}^3C_{W}^{10}}\frac{m_{\mu}^{5}}{M_{Z'}^{4}}k^2.$$
Where
$C^{2}_{W}=cos^{2}\theta_{W}$, $\theta_{W}$ is the Weinberg angle,
$M_{Z'}$ is the mass of the non-universal $U(1)$ gauge boson $Z'$
predicted by TC2 models. In above equations, we have assumed
$m_{\mu}\approx0$, $m_{e}\approx0$ for the processes
$\tau\longrightarrow l_{j}\nu_{l}\bar{\nu_{l}}$ and
$m_{e}\approx0$ for the processes $\mu\longrightarrow
e\nu_{l}\bar{\nu_{l}}$. The widths of the processes
$l_{i}\longrightarrow l_{j}\nu_{\mu}\bar{\nu}_{\mu}$ are equal to
those of the processes $l_{i}\longrightarrow
l_{j}\nu_{e}\bar{\nu}_{e}$. This is because the gauge boson $Z'$
only treats the fermions in the third generation differently from
those in the first and second generations and treats the fermions
in the first generation same as those in the second generation.

The corresponding branching ratios can be written as:
$$Br_{1}=Br^{exp}(\tau\longrightarrow e\nu_{e}\bar{\nu}_{\tau})\frac{\Gamma_{1}}
{\Gamma(\tau\longrightarrow e\nu_{e}\bar{\nu}_{\tau})}, \hspace{5mm}
Br_{2}=Br^{exp}(\tau\longrightarrow
e\nu_{e}\bar{\nu}_{\tau})\frac{\Gamma_{2}}{\Gamma(\tau\longrightarrow
e\nu_{e}\bar{\nu}_{\tau})}, $$
$$Br_{3}=\frac{\Gamma_{3}}{\Gamma(\mu\longrightarrow e\nu_{e}\bar{\nu_{\mu}})},\hspace{5mm}
Br_{4}=\frac{\Gamma_{4}}{\Gamma(\mu\longrightarrow e\nu_{e}\bar{\nu_{\mu}})},$$
with
$$\Gamma(\tau\longrightarrow e\nu_{e}\bar{\nu}_{\tau})=\frac{m_{\tau}^{5}G_{F}^{2}}{192\pi^{3}},
\hspace{5mm}\Gamma(\mu\longrightarrow
e\nu_{e}\bar{\nu_{\mu}})=\frac{m_{\mu}^{5}G_{F}^{2}}{192\pi^{3}}.$$
Here the Fermi coupling constant $G_{F}=1.16639\times
10^{-5}GeV^{-2}$ and the branching ratio
$Br^{exp}(\tau\longrightarrow
e\nu_{e}\bar{\nu}_{\tau})=(17.83\pm0.06)\%$ \cite{y10}.

To obtain numerical results, we take the SM parameters as
$C_{W}^{2}=0.7685$, $\alpha_{e}=\frac{1}{128.8}$,
$m_{\tau}=1.78GeV$, $m_{\mu}=0.106GeV$ \cite{y10}. It has been
shown that vacuum tilting and the constraints from Z-pole physics
and $U(1)$ triviality require $k_{1}\leq1$ \cite{y11}. The limits
on the $Z'$ mass $M_{Z'}$ can be obtained via studying its effects
on various experimental observables \cite{y9}. For example,
Ref.\cite{y12} has been shown that to fit the electroweak
measurement data, the $Z'$ mass $M_{Z'}$ must be larger than
$1TeV$. As numerical estimation, we take the $M_{Z'}$ and $k_{1}$
as free parameters.

The branching ratios $Br_{1}$ and $Br_{2}$ are ploted in Fig.2 and
Fig.3 as functions of $M_{Z'}$ for $k=\lambda=0.22$ ($\lambda$ is
the Wolfenstein parameter  \cite{y13}) and three values of the
parameter $k_{1}$: $k_{1}=0.2$(solid line), 0.5(dotted line),
0.8(dashed line). One can see that the value of $Br_{1}$ is larger
than that of $Br_{2}$ in all of the parameter space of TC2 models.
This is because the extra $U(1)$ gauge boson $Z'$ couple
preferentially to the third generation fermions. The value of the
branching ratio $Br_{1}$ increases from $3.09\times10^{-8}$ to
$7.91\times10^{-6}$ as $M_{Z'}$ decreasing from $4TeV$ to $1TeV$
for $k_{1}=0.5$ and the value of branching ratio $Br_{2}$
increases from $1.26\times10^{-11}$ to $3.23\times10^{-9}$. For
$k_{1}=1$, $M_{Z'}=1TeV$, the branching ratio $Br_{1}$ can reach
$1.6\times10^{-5}$. Certainly, the numerical results are changed
by the value of the flavor mixing parameter $k$. If we take the
maximum values of the parameters  i.e. $(k_{1})_{max}=1$ and
$(k)_{max}=1/\sqrt{2}$, then we have $Br_{1}=1.63\times10^{-4}$,
$Br_{2}=1.67\times10^{-8}$ and $Br_{1}=1.02\times10^{-5}$,
$Br_{2}=1.04\times10^{-9}$ for $M_{Z'}=1TeV$ and $2TeV$,
respectively.

\begin{figure}[h]
\begin{center}
\begin{picture}(200,210)(0,0)
\put(-50,0){\epsfxsize 90mm \epsfbox{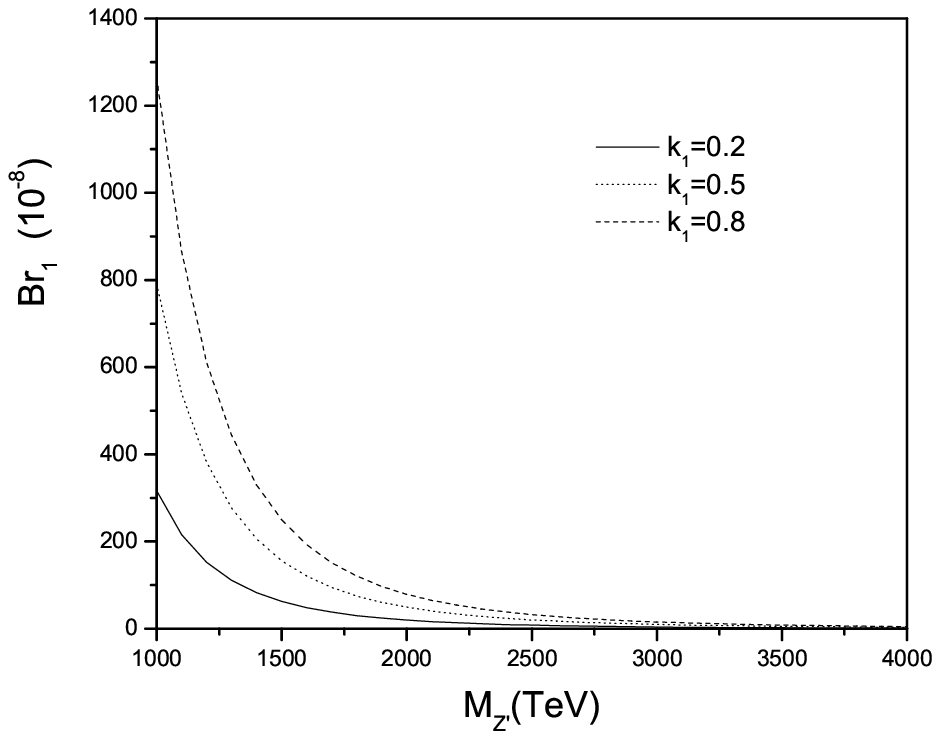}}
 \put(-105,-5){Fig.2: Branching ratio $Br_{1}$ as a function of the $Z'$ mass $M_{Z'}$ for the
            flavor mixing}
 \put(-85,-25){factor $k=0.22$ and $k_{1}=0.2$(solid line), 0.5(dotted line), 0.8(dashed line).}
\end{picture}
\end{center}
\end{figure}
\vspace{1cm}

The extra $U(1)$ gauge boson $Z'$ can also contribute to the LFV
process $\mu\longrightarrow 3e$. The relevant decay width arisen
from the $Z'$ exchange can be written as:
$$\Gamma(\mu\longrightarrow 3e)=\frac{25\alpha_{e}^{5}}{384\pi k_{1}^{3}
C_{W}^{10}}\frac{m_{\mu}^{5}}{M_{Z'}^{4}}k^{2}.$$
The current experimental upper limit is
$Br^{exp}(\mu\longrightarrow 3e)\leq 1\times10^{-12}$ \cite{y14}.
Therefore, the present experimental bound on the LFV process
$\mu\longrightarrow 3e$ can give severe constraints on  the free
parameters of TC2 models. Then the branching ratios $Br_{1}$,
$Br_{2}$, $Br_{3}$ and $Br_{4}$ can be written as:
$$Br_{1}\leq
\frac{k_{1}^{4}C_{W}^{8}}{5\alpha_{e}^{4}}Br^{exp}(\tau\longrightarrow
e\nu_{e}\bar{\nu_{\tau}})Br^{exp}(\mu\longrightarrow3e),$$
$$Br_{2}\leq \frac{k_{1}^{2}C_{W}^{4}}{5\alpha_{e}^{2}}Br^{exp}(\tau\longrightarrow
e\nu_{e}\bar{\nu_{\tau}})Br^{exp}(\mu\longrightarrow3e),$$
$$Br_{3}\leq \frac{k_{1}^{2}C_{W}^{4}}{5\alpha_{e}^{2}}Br^{exp}(\mu\longrightarrow3e),$$
$$Br_{4}\leq \frac{1}{5}Br^{exp}(\mu\longrightarrow3e).$$
Observably, the maximum values of these branching ratios  are only
dependent on the free parameter $k_{1}$. For $k_{1}\leq1$, we have
$Br_{1}\leq3.42\times10^{-6}$, $Br_{2}\leq3.49\times10^{-10}$,
$Br_{3}\leq1.96\times10^{-9}$ and $Br_{4}\leq2\times10^{-13}$.
\begin{figure}[h]
\begin{center}
\begin{picture}(200,100)(0,0)
\put(-50,-120){\epsfxsize 90mm \epsfbox{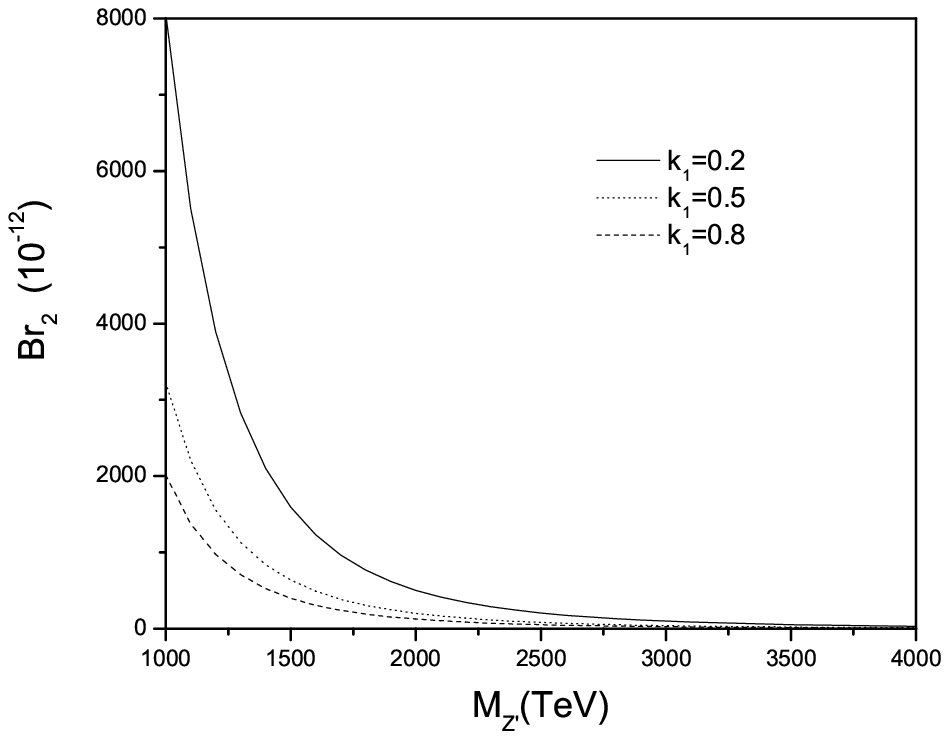}}
\put(-30,-130){Fig.3: Same as Fig.2 but for $Br_{2}$. }
\end{picture}
\end{center}
\end{figure}
\vspace{4.5cm}

Extra gauge bosons $Z'$ are the best motivated extensions of the
SM. If discovered they would represent irrefutable proof of new
physics, most likely that the SM gauge groups must be extended
\cite{y15}. If these extensions are associated with flavor
symmetry breaking, the gauge interactions will not be
flavor-universal \cite{y12}, which predict the existence of
non-universal gauge bosons $Z'$. After the mass diagonalization
from the flavor eigenbasis into the mass eigenbasis, the
non-universal gauge interactions result in the tree-level FC
couplings. Thus, the $Z'$ may have significant contributions to
some FCNC processes. In this letter, we study the contributions of
the non-universal gauge bosons $Z'$ predicted by TC2 models to the
LFV processes $l_{i}\longrightarrow l_{j}\nu_{l}\bar{\nu}_{l}$. We
find that the branching ratios $B_{r}(\tau\longrightarrow
l_{j}\nu_{\tau}\bar{\nu}_{\tau})$ are larger than the branching
ratios $B_{r}(\tau\longrightarrow l_{j}\nu_{l}\bar{\nu}_{l})
(l=\mu$ or $e)$ in all of the parameter space. Over a wide range
of parameter space, we have $B_{r}(\tau\longrightarrow
l_{j}\nu_{\tau}\bar{\nu}_{\tau})\sim 10^{-6}$ and
$B_{r}(\tau\longrightarrow l_{j}\nu_{l}\bar{\nu}_{l})\sim 10^{-9}
$. For $k_{1}=1$, $M_{Z'}=1TeV$ and $k=1/\sqrt{2}$, the value of
the branching ratio $B_{r}(\tau\longrightarrow
l_{j}\nu_{\tau}\bar{\nu}_{\tau})$ can reach $1.63\times10^{-4}$.
Considering the bounds given by the experimental upper limit
$Br^{exp}(\mu\longrightarrow3e)\leq1\times10^{-12}$ on the free
parameters of TC2 models, we further give the upper limits of the
LFV processes $l_{i}\longrightarrow l_{j}\nu_{l}\bar{\nu}_{l}$.
The  results are $Br({\tau\longrightarrow
l_{j}\nu_{\tau}\bar{\nu}_{\tau}})\leq3.42\times10^{-6}$,
$Br({\tau\longrightarrow
l_{j}\nu_{l}\bar{\nu}_{l}})\leq3.49\times10^{-10}$,
$Br({\mu\longrightarrow
e\nu_{\tau}\bar{\nu}_{\tau}})\leq1.96\times10^{-9}$ and
$Br({\mu\longrightarrow
e\nu_{l}\bar{\nu}_{l}})\leq2\times10^{-13}$ $(l=\mu$ or $e)$. We
hope that the results may be useful to partly explain the data
neutrino oscillations. The future neutrino experiment data might
be used to test TC2 models.

\vspace{1.5cm} \noindent{\bf Acknowledgments}

Chongxing Yue would like to thank Professor J. M. Yang for useful
discussions. This work was supported  by the National Natural
Science Foundation of China (90203005).

\end{document}